\documentclass[aps,prl,twocolumn,superscriptaddress,floatfix,preprintnumbers]{revtex4-2}

\usepackage[utf8]{inputenc}
\usepackage{amsmath}
\usepackage{amssymb}

\usepackage{enumerate}
\usepackage{tikz}
\usepackage{feynmf}
\usepackage{slashed}
\usepackage{braket}
\usepackage[toc,page]{appendix}
\usepackage{url}
\usepackage{graphicx}
\usepackage[pdftex,bookmarks,linktocpage,pdfpagelabels,plainpages=false,hyperfigures,linkcolor=blue,citecolor=blue]{hyperref} %for hypertext links
\hypersetup{colorlinks=true}

\usepackage{mathrsfs,amssymb}  
\usepackage{cancel}
\usepackage[normalem]{ulem}

\usepackage{tikz}
\usetikzlibrary{decorations.markings}
\usetikzlibrary{positioning}
\usetikzlibrary{shapes}
\usetikzlibrary{calc}

\newcommand{\beq}{\begin{equation}}
\newcommand{\eeq}{\end{equation}}

\newcommand{\fref}[1]{\hyperref[#1]{FIG.~\ref*{#1}}}
\newcommand{\Fref}[1]{\hyperref[#1]{FIG.~\ref*{#1}}}

\begin{document}

\author{Doojin~Kim}
\email{doojin.kim@tamu.edu}
\affiliation{Mitchell Institute for Fundamental Physics and Astronomy,
Department of Physics and Astronomy, Texas A\&M University, College Station, TX 77843, USA}

\author{Kyoungchul Kong}
\email{kckong@ku.edu}
\affiliation{Department of Physics and Astronomy, University of Kansas, Lawrence, KS 66045, USA}

\author{Konstantin T. Matchev}
\email{matchev@ufl.edu}
\affiliation{Institute for Fundamental Theory, Physics Department, University of Florida, Gainesville, FL 32611, USA}

\author{Myeonghun Park}
\email{parc.seoultech@seoultech.ac.kr}
\affiliation{Institute of Convergence Fundamental Studies, Seoultech, Seoul, 01811, Korea}
\affiliation{School of Physics, KIAS, Seoul 02455, Korea}
\affiliation{Center for Theoretical Physics of the Universe, Institute for Basic Science, Daejeon 34126 Korea}

\author{Prasanth Shyamsundar}
\email{prasanth@fnal.gov}
\affiliation{Fermilab Quantum Institute, Fermi National Accelerator Laboratory, Batavia, IL 60510, USA}

\preprint{FERMILAB-PUB-21-244-QIS}
\preprint{MI-TH-2111}

\title{Deep-Learned Event Variables for Collider Phenomenology}

\begin{abstract}
The choice of optimal event variables is crucial for achieving the maximal sensitivity of experimental analyses. Over time, physicists have derived suitable kinematic variables for many typical event topologies in collider physics. Here we introduce a deep learning technique to design good event variables, which are sensitive over a wide range of values for the unknown model parameters. We demonstrate that the neural networks trained with our technique on some simple event topologies are able to reproduce standard event variables like invariant mass, transverse mass, and stransverse mass. The method is automatable, completely general, and can be used to derive sensitive, previously unknown, event variables for other, more complex event topologies.
\end{abstract}

\maketitle

\noindent {\bf Introduction.} Data in collider physics is very high-dimensional, which brings a number of challenges for the analysis, encapsulated in ``the curse of dimensionality"\,\cite{Bellman}. Mapping the raw data to reconstructed objects involves initial dimensionality reduction in several stages, including track reconstruction, calorimeter clustering, jet reconstruction, etc. Subsequently, the kinematics of the reconstructed objects is used to define suitable analysis variables, adapted to the specific channel and targeted event topology. Each such step is essentially a human-engineered feature-extraction process from complicated data to a handful of physically meaningful quantities. While some information loss is unavoidable, physics principles and symmetries help keep it to a minimum.

In this letter, we shall focus on the last stage of this dimensionality reduction chain, namely, the optimal construction of kinematic variables, which is essential to expedite the discovery of new physics and/or to improve the precision of parameter measurements. By now, the experimentalist's toolbox contains a large number of kinematic variables, which have been thoroughly tested in analyses with real data (see \cite{Han:2005mu,Barr:2010zj,Barr:2011xt,Matchev:2019sqa} for reviews). The latest important addition to this set are the so-called ``singularity variables" \cite{Kim:2009si,Rujula:2011qn,DeRujula:2012ns,Kim:2019prx,Matchev:2019bon}, which are applicable to missing energy events --- the harbingers of dark matter production at colliders. In the machine learning era, a myriad of algorithms have been invented or adopted to tackle various tasks that arise in the analysis of collider data, e.g., signal--background discrimination (see \cite{Feickert:2021ajf} for a continuously updated complete review of the literature). \emph{Under the hood}, the machines trained in these techniques could learn to construct useful features from the low-level event description, because they are relevant to the task at hand. But it is difficult to interpret what exactly the machines have learned in the process \cite{Chang:2017kvc,Faucett:2020vbu}. Furthermore, it is rarely studied whether the human-engineered features are indeed the best event variables for certain purposes, and whether machines can outperform theorists at constructing event variables.

These two issues, explainability and optimality, are precisely the two questions which we shall address in this letter. We shall introduce a new technique for training neural networks to directly \emph{output} useful features or event variables (which offer sensitivity over a range of unknown parameter values). This allows for explainability of the machine's output by comparison against known features in the data. At the same time, it is important to verify that the variables obtained using our technique are indeed the optimal choice, and we will test this by directly comparing them against the human-engineered variables that are known to be optimal for their respective event topologies. Once we have validated our training procedure in this way, we could extend it to more complex event topologies and derive novel kinematic variables in interesting and difficult scenarios.

Understanding how and what a neural network (NN) learns is a difficult task. Here we shall consider relatively simple physics examples that are nevertheless highly non-trivial from a machine learning point of view: (1) visible two-body decay (to two visible daughter particles); (2) semi-invisible two-body decay (to one visible and one invisible daughter particle); (3) semi-invisible two-body decays of pair-produced particles. It is known that the relevant variables in those situations are the invariant mass $m$, the transverse mass $m_T$ \cite{Barger:1983wf,Smith:1983aa} and the stransverse mass $m_{T2}$ \cite{Lester:1999tx}, respectively. We will demonstrate that in each case, the NN can be trained to learn the corresponding physics variable in the reduced latent space. The method can be readily generalized to more complex cases to derive deep-learned, next-generation event variables.

\vskip 2mm
\noindent\textbf{Methodology.} Let $X$ represent the high-dimensional input features from a collision event, e.g., the 4-momenta of the reconstructed physics objects. Let $V(X)$ be a low-dimensional event variable constructed from $X$. In this work, we shall model the function $V$ using a neural network, where for notational convenience, the dependence of $V$ on the architecture and weights of the network will not be explicitly indicated. We imagine that $V$ retains the relevant physics information and will be the centerpiece of an experimental study of a theory model with a set of unknown parameters $\Theta$. The goal is to train the NN encoding the function $V$ to be ``useful'' over a wide range of values for $\Theta$. For this purpose, we will need to train with events generated from a range of $\Theta$ values. Note that this is a departure from the traditional approach in particle physics, where training is done for specific study points with fixed values of $\Theta$. In addition, we will have to quantify the usefulness of a given event variable $V(X)$, as explained in the remainder of this section.

\vskip 2mm
\noindent\textit{Intuition from Information Theory.}
Each event $X$ carries some information about the underlying model parameter values from which it was produced. Some of this information could be lost when reducing the dimensionality of the data from $d_X\equiv \dim(X)$ to $d_V\equiv \dim(V)$, as a consequence of the data processing inequality \cite{DPI}. Good event variables minimize this information loss, and efficiently retain the information about the underlying parameter values $\Theta$ \cite{Tishby:2000,Iten2020}. This is precisely why the invariant mass $m$, the transverse mass $m_T$ and the stransverse mass $m_{T2}$ have been widely used in particle physics for mass parameter measurements and for new physics searches.

The mutual information of $V$ and $\Theta$ is given by
\begin{equation}
 I(V;\Theta) \equiv \int\limits_{\mathcal{V}} dv \int\limits_{\Omega} d\theta~p_{(V,\Theta)}(v, \theta)\,\ln\!{\left[\frac{p_{(V,\Theta)}(v, \theta)}{p_V(v)\,p_\Theta(\theta)}\right]}, \label{eq:I_V_Theta}
\end{equation}
where $p_V$ and $p_\Theta$ are the probability distribution functions of $V$ and $\Theta$, respectively, and $p_{(V,\Theta)}$ is their joint distribution. $\mathcal{V}$ and $\Omega$ are the domains of $V$ and $\Theta$, respectively. One can think of $p_\Theta$ as the prior distribution of $\Theta$.\footnote{For convenience, we will adopt the Bayesian interpretation of probability in the presentation of this work.} The distributions $p_{(V,\Theta)}$ and $p_V$ can then be derived from $p_\Theta$ and the conditional distribution $p_{V|\Theta}(v|\theta)$.

The mutual information $I(V;\Theta)$ quantifies the amount of information contained in $V$ about $\Theta$. Therefore, a good event variable $V$ should have relatively high values of $I(V;\Theta)$. From Eq.\,\eqref{eq:I_V_Theta}, one can see that $I(V;\Theta)$ is nothing but the Kullback--Leibler (KL) divergence from (a) the factorized distribution $p_V\otimes p_\Theta$ to (b) the joint distribution $p_{(V,\Theta)}$. The KL divergence, in turn, is a measure of how distinguishable the two distributions (a) and (b) are.

\begin{figure*}
 \centering
 \includegraphics[width=0.85\textwidth]{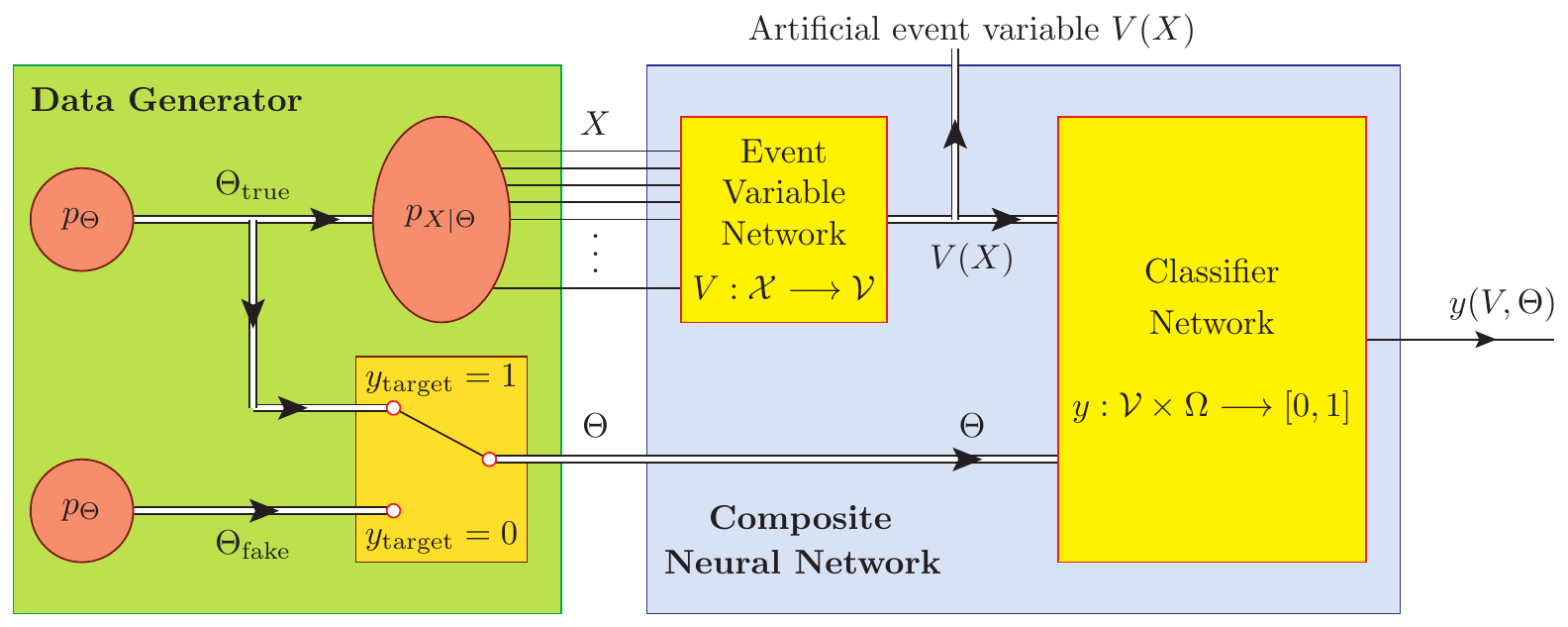}
 \caption{A schematic diagram of the training strategy for the artificial event variable $V$. The left (green) and right (blue) blocks depict the generation of training data and the composite neural network layout, respectively.}
 \label{fig:metanetwork}
\end{figure*}

These observations lead to the following strategy schematically outlined in~\fref{fig:metanetwork}: train the event variable network so that the distributions $p_V\otimes p_\Theta$ and $p_{(V,\Theta)}$ are highly distinguishable. An auxiliary classifier network can then be used for evaluating the distinguishability of the two distributions. The basic blueprint of our training technique will be described next.

\vskip 2mm 
\noindent\textit{Training Data Generation.} 
In order to generate the training data, we start with the two distributions $p_\Theta$ and $p_{X|\Theta}$.  The specific choice of a prior distribution $p_\Theta$ is not crucial --- as long as it allows to sample $\theta$ over a sufficiently wide range (the one in which we want the event variable $V$ to be sensitive) any function will do, and one is further free to impose theoretical prejudice like fine tuning, etc.

$p_{X|\Theta}$ is the distribution of the event $X$ conditional on $\Theta$. General purpose event generators can be used to sample from this distribution. The overall distribution of $X$, namely $p_X$, is given by
\begin{equation}
 p_X(x) = \int\limits_\Omega d\theta~p_\Theta(\theta)~p_{X|\Theta}(x | \theta).
\end{equation}

Our training data consists of two classes, whose generation is illustrated in the left (green) block of \fref{fig:metanetwork}. Each training data point is given by a 2-tuple $(X, \Theta)$ along with the class label $y_\mathrm{target}\in\{0, 1\}$ of the data point. Under class 0, $X$ and $\Theta$ are independent of each other and their joint distribution is given by $p_X\otimes p_\Theta$. This is accomplished by simply replacing the true value of $\Theta$ used to generate $X$ with a fake one for the datapoints in class 0. Under class 1, the joint distribution of $(X, \Theta)$ is given by
\begin{equation}
 p_{(X,\Theta)}(x, \theta) = p_{X|\Theta}(x|\theta)~p_\Theta(\theta).
\end{equation}

\vskip 2mm
\noindent\textit{Event Variable Training.} As shown in the right (blue) block \fref{fig:metanetwork}, we then set up a composite network for classifying the data points $(X,\Theta)$ into the two classes. The composite network consists of two parts. First, an event variable network (EVN) takes the high-dimensional collider event information $X$ as input and returns a low-dimensional $V(X)$ as output. As indicated, this network parameterizes the artificial event variable function $V(X)$, which is precisely what we are interested in training. The output layer of the EVN network does {\em not} use an activation function (or, equivalently, uses the identity activation). Since $d_V\ll d_X$, the main task of the EVN network is to perform the needed dimensionality reduction.  However, to ensure that this retains the maximal amount of information, we introduce an auxiliary classifier network which takes the event variable $V(X)$ and the parameters $\Theta$ as input and returns a one-dimensional output, $y(V, \Theta) \in [0,1]$. Note that the input received by the auxiliary network is distributed as $p_V\otimes p_\Theta$ under class 0, and as $p_{(V,\Theta)}$ under class 1.

The information bottleneck \cite{Tishby:2000} $V(X)$ created by the EVN module is optimized by simply training the composite network as a classifier for the input data $(X,\Theta)$, using the class labels $y_\mathrm{target}$ as the supervisory signal.

\vskip 2mm
\noindent\textbf{Experiments.} The EVN module in the network architecture from \fref{fig:metanetwork} reduces the original $d_X$-dimensional features to a $d_V$-dimensional subspace of event variables, which by construction are guaranteed to be highly sensitive to the theory model parameters $\Theta$, but without any explicit dependence on them. Such variables have been greatly valued in collider phenomenology, and a significant number have been proposed and used in experimental analyses. As a proof of principle, we shall now demonstrate how our approach is able to reproduce the known kinematic variables in a few simple but non-trivial examples. Here we shall only consider one variable at a time, i.e., $d_V=1$, postponing the case of $d_V>1$ to future work \cite{KKMPS}. 

\vskip 2mm 
\noindent\textit{Example 1: Fully visible two-body decay.} First we consider the fully visible decay of a parent particle $A$ into two massless visible daughter particles, $A\rightarrow b\,c$. The parameter $\Theta$ in this example is the mass $m_A$ of the mother particle $A$. The event $X$ is specified by the 4-momenta of the daughter particles $p_b$ and $p_c$, leading to $d_X = 8$.

The prior $p_\Theta$ is chosen to sample $m_A$ uniformly in the range $[100, 500]$ GeV. For each sampled value of $m_A$, we generate an event as follows. A generic boost for the parent particle $A$ is obtained by isotropically picking the direction for its momentum and uniformly sampling its lab frame energy in the range $[m_A, 1500\text{ GeV}]$. Subsequently, $A$ is decayed on-shell into two massless particles (isotropically in its own rest frame), so that the input data consists of the lab-frame final state 4-momenta $\{p_b,p_c\}\equiv\{E_b, \vec{p}_{b}, E_c, \vec{p}_{c}\}$.

All the neural networks used in this work were implemented in \textsc{TensorFlow} \cite{tensorflow}. For the event variable network, we used a sequential fully connected architecture with 5 hidden layers. The hidden layers, in order, have 128, 64, 64, 64, and 32 nodes, all of them using ReLU as the activation function. The output layer has one node with no activation function. The classifier network is a fully connected network with 3 hidden layers (16 nodes each, ReLU activation). The output layer has one node with sigmoid activation. These two networks were combined as shown in the right (blue) block of \fref{fig:metanetwork} and trained with 2.5 million events total (50--50 split between classes 0 and 1), out of which 20\% was set aside for validation. The network was trained for 20 epochs with a mini-batch size of 50, using the Adam optimizer and the binary crossentropy loss function.

\begin{figure*}[t]
\begin{center}
\includegraphics[width=.32\textwidth]{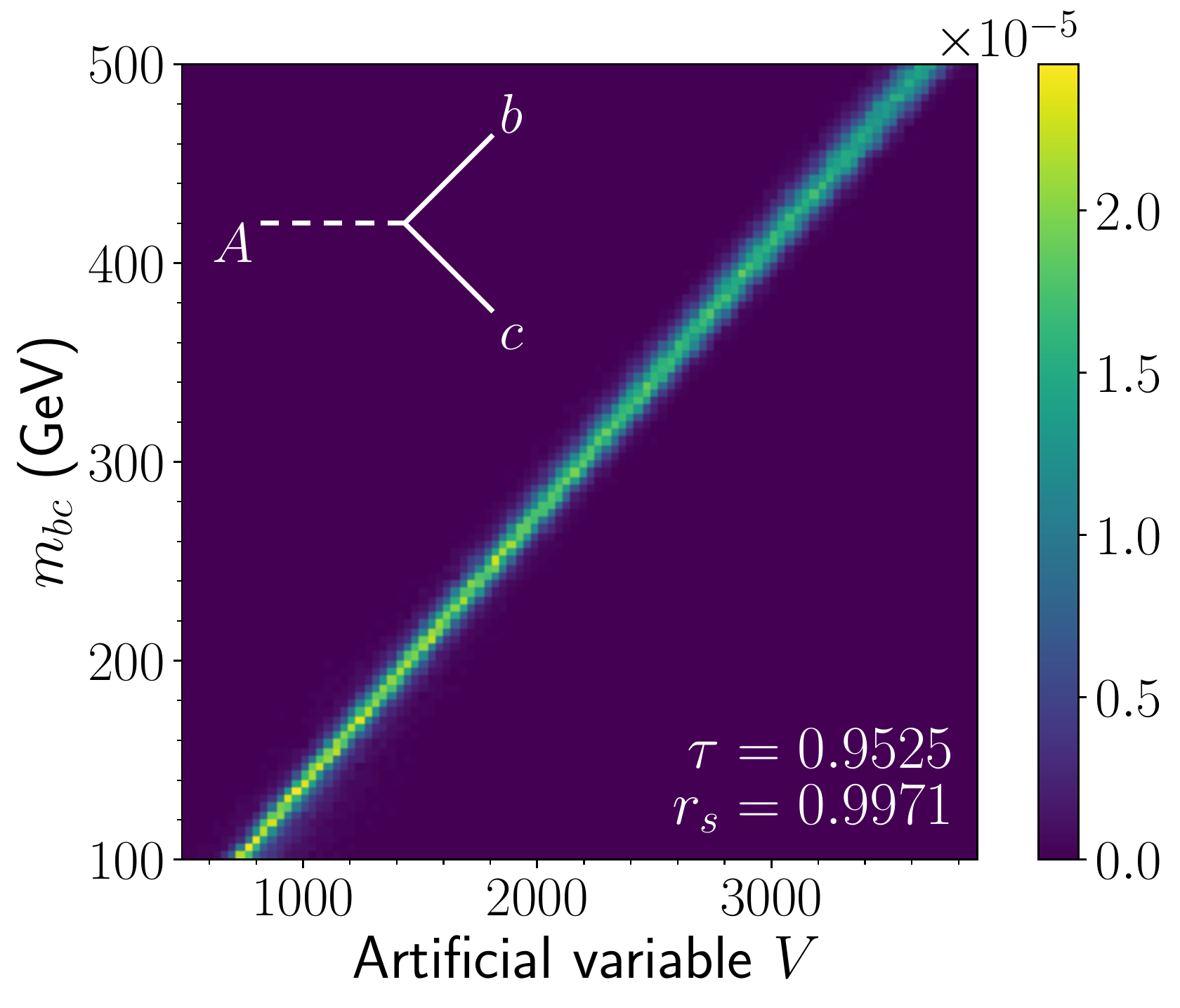}~
\includegraphics[width=.32\textwidth]{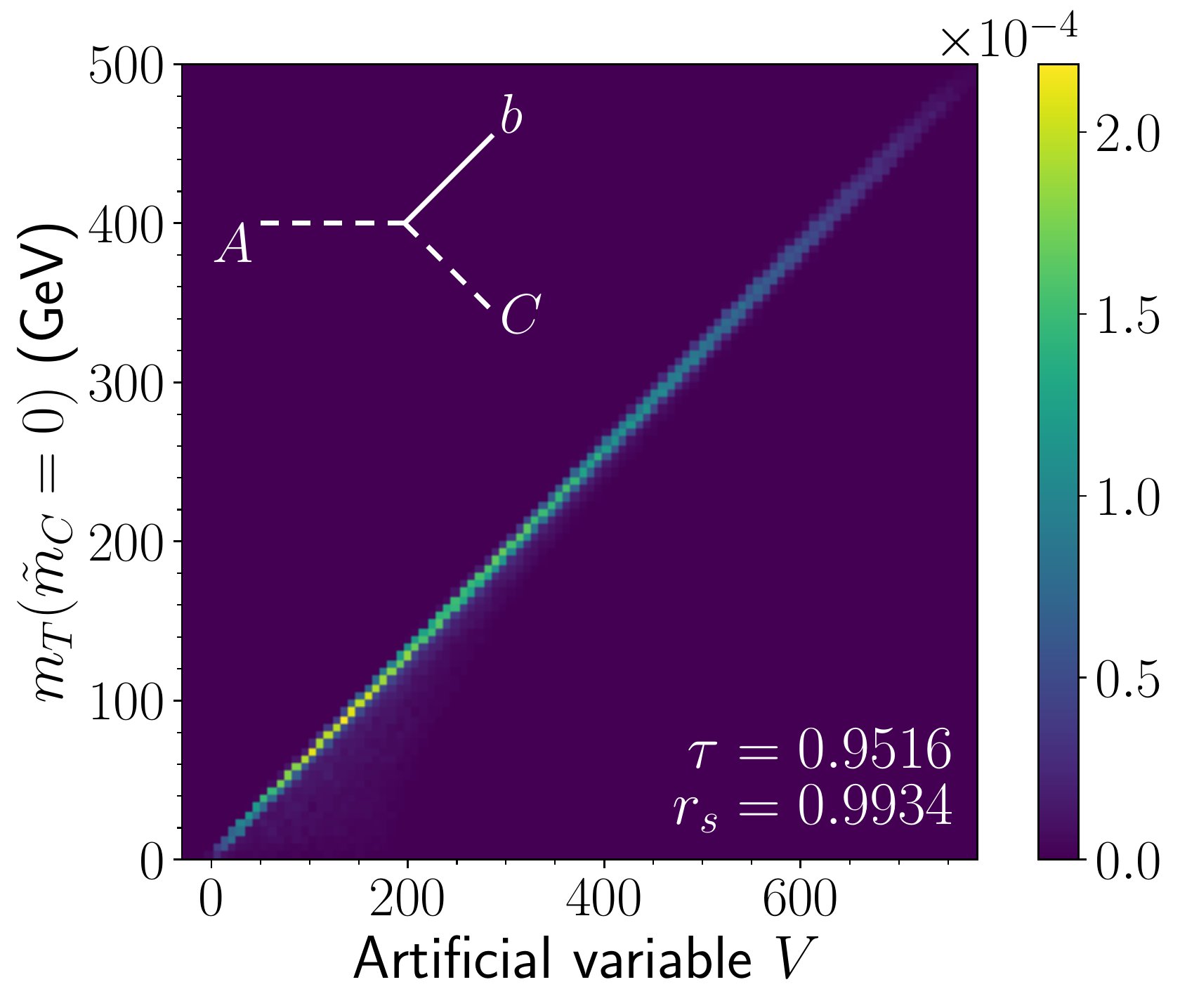}~ 
\includegraphics[width=.32\textwidth]{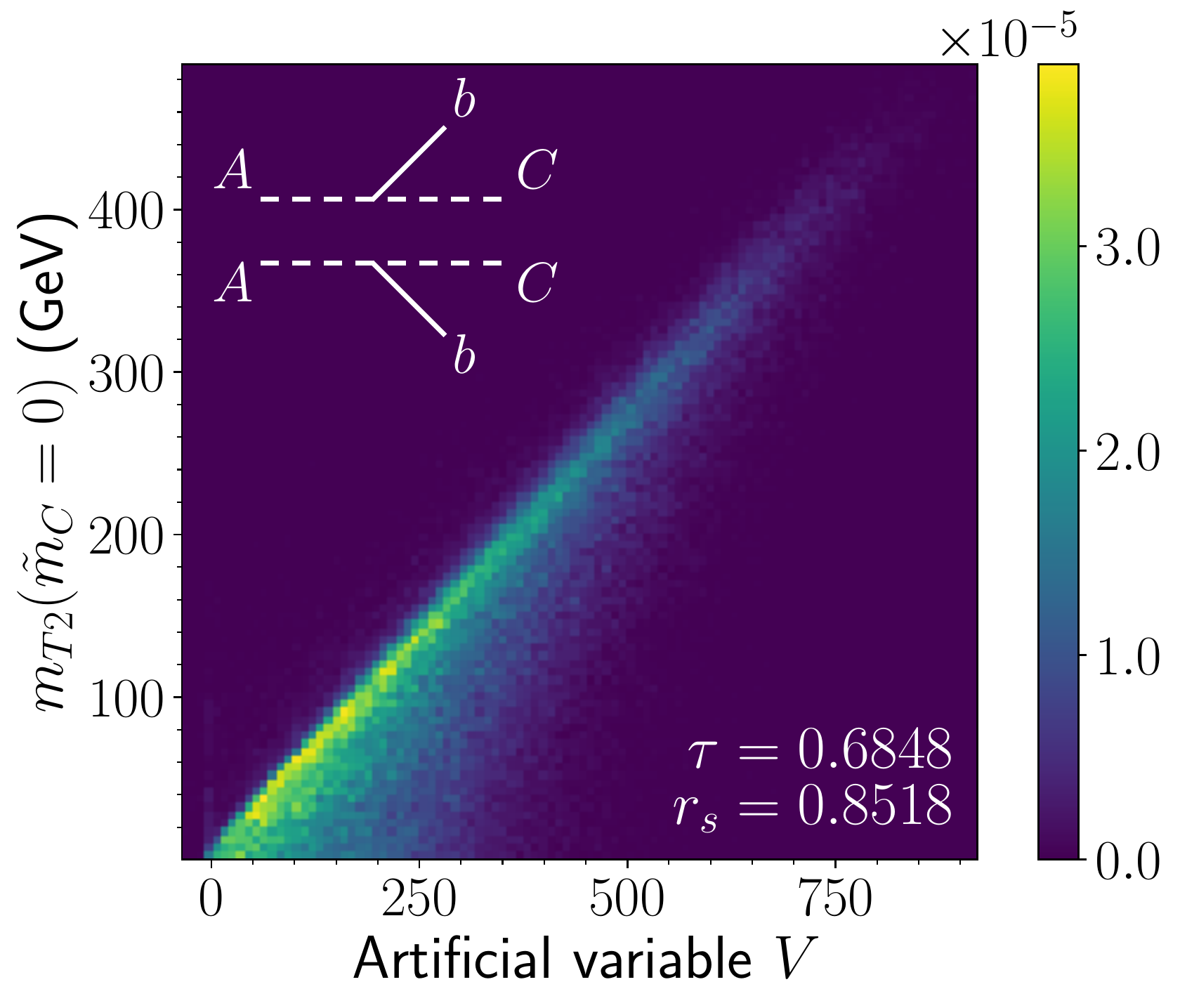} \\
\includegraphics[width=.31\textwidth]{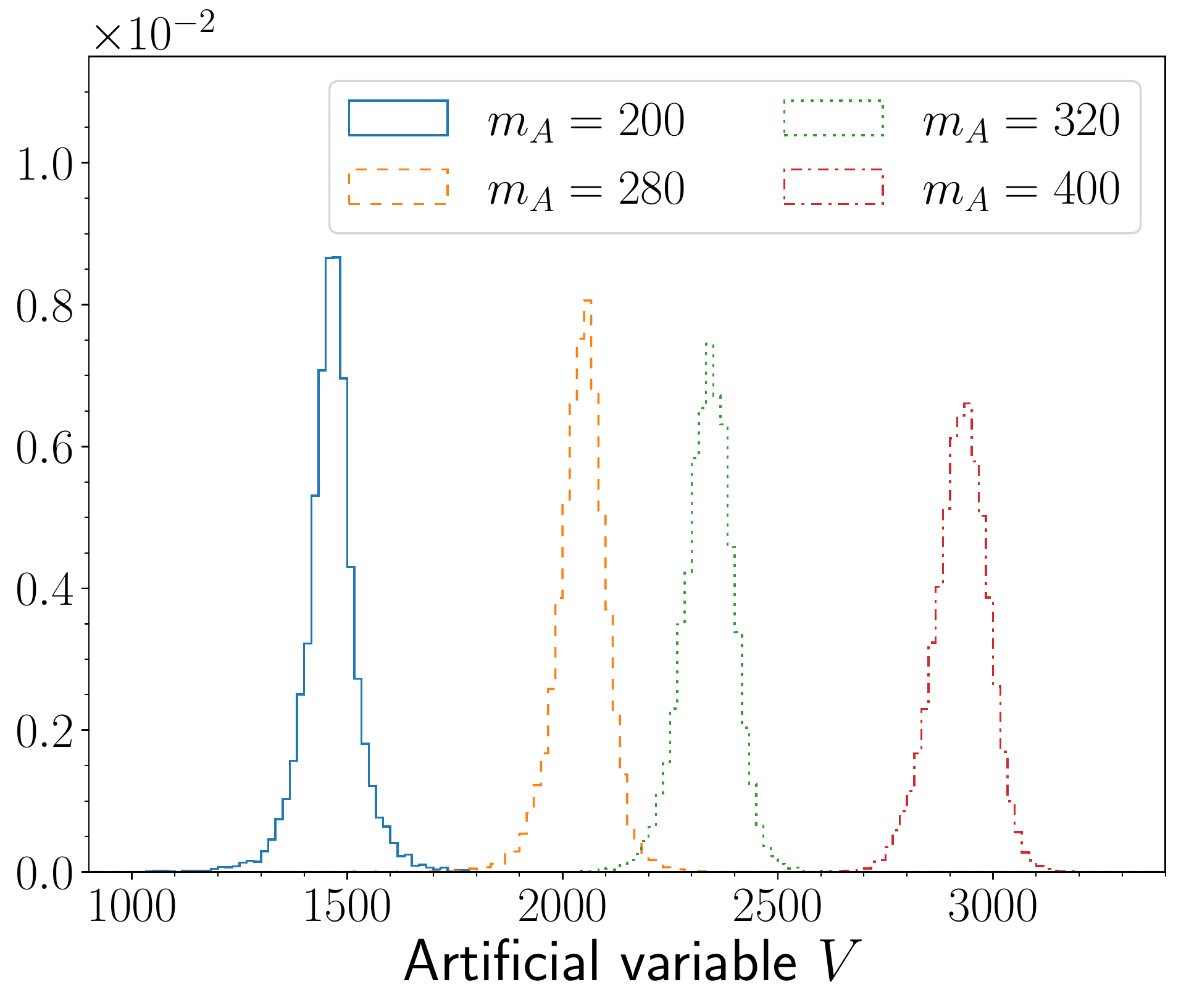}~~~
\includegraphics[width=.31\textwidth]{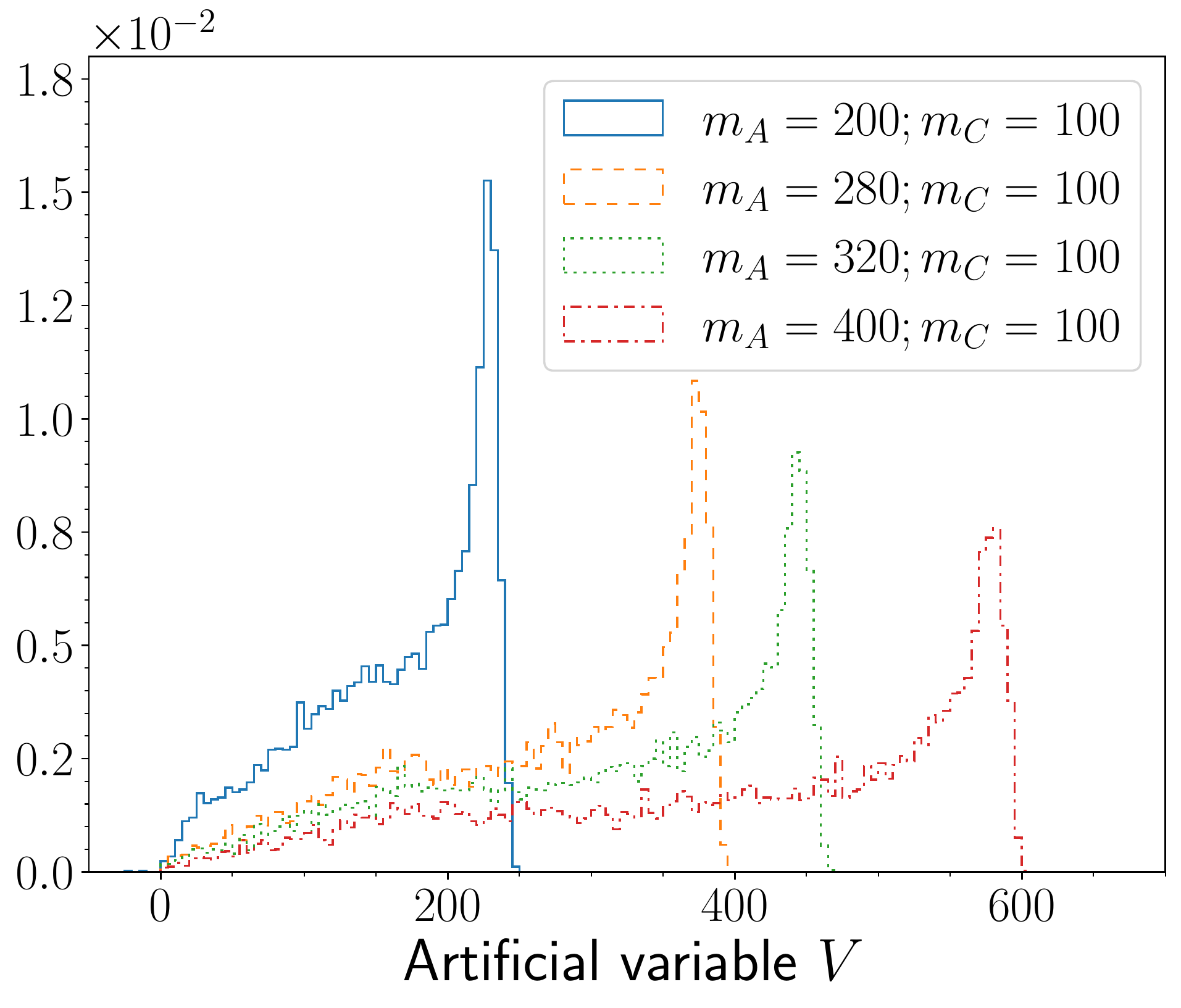}~~~
\includegraphics[width=.31
\textwidth]{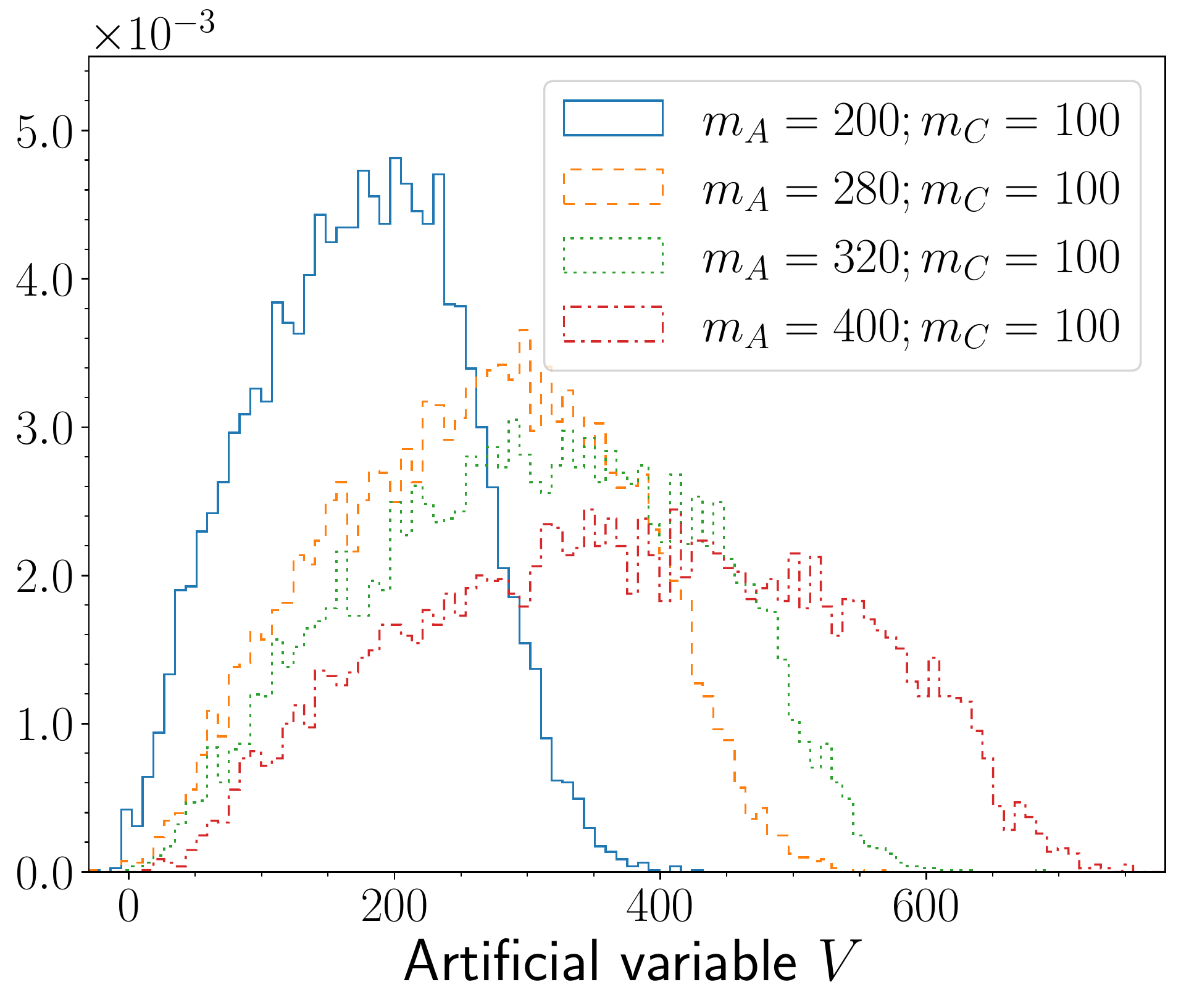}
\end{center}
\caption{\label{fig:example}
Top row: correlation plots between the artificial variable $V$ and a relevant human-engineered variable for each of the three examples considered in the text ($m_{bc}$, $m_T$ and $m_{T2}$ from left to right). Bottom row: unit-normalized distributions of the corresponding artificial variable $V$ for different mass inputs.
}
\end{figure*}

For the event topology considered in this example, it is known that the event variable most sensitive to the value of $m_A$ is the invariant mass of the daughter particles
\begin{equation}
 m_{bc}=\sqrt{(E_b+E_c)^2 - (\vec{p}_{b}+\vec{p}_{c})^2~}\,,
\end{equation}
as well as any variable that is in one-to-one correspondence with it. In order to test whether our artificial event variable $V$ learned by the NN correlates with $m_{bc}$, we show a heatmap of the joint distribution of $(V, m_{bc})$ in the upper-left panel of \fref{fig:example}. Here, and in what follows, the heatmap is generated using a separate test dataset with $10^5$ events. In the plot we also show two nonparametric correlation coefficients, namely Kendall's $\tau$ coefficient \cite{Kendall} and Spearman's rank correlation coefficients $r_s$ \cite{spearman}. A value of $\pm1$ for them would indicate one-to-one correspondence. Our results depict an almost perfect correspondence between $V$ and $m_{bc}$. Here, and in what follows, we append an overall minus sign to $V$ if needed, to make the correlations positive and the plots in \fref{fig:example} intuitive.

In practice, the artificial variable can be used to compare the data against templates simulated for different values of $\Theta$. To illustrate this usage, in the lower-left panel of \fref{fig:example} we show unit-normalized distributions of the deep-learned variable $V$ for several different values of $m_A=\{200,280,320,400\}$ GeV. It is seen that the distributions are highly sensitive to the parameter choice $m_A$ and, if needed, $V$ can be calibrated so that the peak location directly corresponds to $m_A$. The observed spread around the peak values in the histogram, as well as the less than perfect correspondence between $V$ and $m_{bc}$, are due to limitations in the NN architecture and training.

\vskip 2mm 
\noindent\textit{Example 2: Semi-visible two body decay.} Next we consider the semi-visible two-body decay of a particle $A$ into a massless visible particle $b$ and a possibly massive invisible particle $C$, $A \rightarrow b\,C$, where $A$ is singly produced (with zero transverse momentum). The parameter $\Theta$ is two-dimensional: $(m_A, m_C)$. The event $X$ is specified by the 4-momentum $p_b=\{E_b, \vec{p}_{bT}, p_{bz}\}$ of $b$ and the missing transverse momentum, leading to $d_X=6$.

We generate $(m_A, m_C)$ by uniformly sampling $(m_A, \delta_m)$ in the region defined by $100\text{ GeV}\leq m_A\leq 500\text{ GeV}$ and $0\leq \delta_m\leq m_A$, where $\delta_m \equiv (m_A-m_C)^2/m_A$. This choice of prior ensures that the relevant mass difference parameter in this event topology $\mu\equiv(m_A^2-m_C^2)/m_A$ is adequately sampled in the range $[0, 500]$ GeV. For the given value of $m_A$, we generate an event as follows. The parent particle $A$ is boosted along the beam axis $\pm z$ (with equal probability) to an energy chosen uniformly in the range $[m_A, 1500\text{ GeV}]$. The particle $A$ is decayed on-shell into $b$ and $C$, isotropically in its own rest frame. The details of network architectures and training are the same as in Example 1.

The relevant variable for this event topology is the transverse mass $m_T$, which in our setup is given by
\begin{equation}
 m_T(\tilde{m}_C) \equiv p_{bT} + \sqrt{p_{bT}^2 + \tilde{m}_C^2~},
\end{equation}
where the choice of mass ansatz $\tilde{m}_C$ for the mass of the invisible particle $C$ does not affect the rank ordering of the events. For concreteness in what follows we shall use $\tilde{m}_C=0$. The corresponding heatmap of the joint distribution $(V,m_T)$ and unit-normalized distributions of the variable $V$ for several choices of $m_A=\{200,280,320,400\}$ GeV and $m_C=100$ GeV are shown in the middle panels of \fref{fig:example}. Once again, we observe an almost perfect correlation between $V$ and $m_T$, and a high sensitivity of the $V$ distributions to the input masses.

\vskip 2mm 
\noindent\textit{Example 3: Symmetric semi-visible two body decays.} Finally, we consider the exclusive production at a hadron collider of two equal-mass parent particles $A_1$ and $A_2$ which decay semi-visibly as $A_1 A_2 \rightarrow (b_1\,C_1)\,(b_2\,C_2)$. The parameter $\Theta$ is given by $(m_A, m_C)$, and the event $X$ is described by the 4-momemta of $b_1$ and $b_2$, and the missing transverse momentum, leading to  $d_X=10$.

The masses $(m_A, m_C)$ are generated as in Example 2. In order to avoid fine turning the network to the details of a particular collider, we uniformly sampled the invariant mass $m_{A_1A_2}$ of the $A_1A_2$ system in the range $[2m_A, 1500\text{ GeV}]$ and the lab-frame energy of the $A_1A_2$ system in the range $[m_{A_1A_2}, 2500\text{ GeV}]$. The direction of the system was chosen to be along $\pm z$ with equal probability. The direction of $A_1$ is chosen isotropically in the rest frame of the $A_1A_2$ system. $A_1$ and $A_2$ are both decayed on-shell, isotropically in their respective rest frames. The details of network architectures and training are the same as in Example 1.

The straightforward generalization of the idea of the transverse mass to the considered event topology leads to the stransverse mass variable $m_{T2}(\tilde m_C)$ \cite{Lester:1999tx}. In the upper-right panel of \fref{fig:example} we show a heatmap of the joint distribution of $(V, m_{T2}(0))$, which reveals reasonably good, but not perfect correlation, implying that the artificial event variable encapsulates information beyond $m_{T2}$. This could have been expected for the following two reasons: 1) unlike the previous two examples of singular variables with sharp features in their distributions, $m_{T2}$ does not belong to the class of singular variables \cite{Matchev:2019bon}; 2) $m_{T2}$ only uses a subset of the available kinematic information, namely the transverse momentum components. In contrast, the artificial kinematic variable can use all of the available information, and in a more optimal way. The lower-right panel of \fref{fig:example} displays unit-normalized distributions of the artificial variable for several choices of $m_A$ and fixed $m_C=100$ GeV, again demonstrating the sensitivity of $V$ to the mass spectrum.

\vskip 2mm
\noindent\textbf{Discussion and outlook.} We proposed a new deep learning technique pictorially summarized in \fref{fig:metanetwork} which allows the construction of event variables from a set of training data produced from a given event topology. The novel component is the simultaneous training for varying parameters $\Theta$ which allows the algorithm to capture the underlying phase space structure irrespective of the benchmark study point. This is the first such method for constructing event variables with neural networks and can be applied to other, more challenging event topologies in particle physics and beyond. In future applications of the method one could enlarge the dimensionality of the latent space to $d_V>1$ and supplement the training data with additional features like tagging and timing information, etc. By manipulating the specifics of the generation of the training data, one can control what underlying physics effects are available for the machine to learn from, and what physical parameters the machine-learned variable will be sensitive to. Our method opens the door to new investigations on intepretability and explainability by incorporating modern representation learning approaches like contrastive learning \cite{Le-Khac2020}.

\vskip 2mm
\noindent\textbf{Code and Data Availability.}
The code and data that support the findings of this study are openly available at the following URL: \url{https://gitlab.com/prasanthcakewalk/code-and-data-availability/} under the directory named \texttt{arXiv\_2105.xxxxx}.

% % % % % Acknowledgements % % % % %
\section*{Acknowledgements}
We are indebted to the late Luc Pape for great insights and inspiration. This work is supported in parts by US DOE DE-SC0021447 and DE-FG02-13ER41976. 
MP is supported by Basic Science Research Program through the National Research Foundation of Korea Research Grant No.\,NRF-2021R1A2C4002551.
PS is partially supported by the U.S. Department of Energy, Office of Science,
Office of High Energy Physics QuantISED program under the grants “HEP Machine Learning
and Optimization Go Quantum”, Award Number 0000240323, and “DOE QuantiSED Consortium
QCCFP-QMLQCF”, Award Number DE-SC0019219.

This manuscript has been authored by Fermi Research Alliance, LLC under Contract No. DEAC02-07CH11359 with the U.S. Department of Energy, Office of Science, Office of High Energy
Physics.

\end{document}